\documentclass[journal=jacsat,manuscript=article]{achemso}

\usepackage[version=3]{mhchem} 
\usepackage{subfigure}

\author{Vladimir Lankevich}
\author{Eric Bittner}
\affiliation{~Department of Chemistry, University of Houston, 4800 Calhoun St., Houston, TX,
USA.}
\email{bittner@uh.edu}
\phone{+1 832-316-8849}
\fax{+1 713-743-2709}

\title[An \textsf{achemso} demo]
  {Effect of Disorder on Free Energy and Open-Circuit Voltage of Organic Photovoltaic Systems}

\abbreviations{IR,NMR,UV}

\begin{document}

\begin{abstract}
Organic Photovoltaic devices (OPVs) are becoming adequately cost and energy efficient to be considered a good investment
and it is, therefore, especially important to have a concrete understanding of their operation. We compute energies of charge-transfer (CT) states of the model donor-acceptor lattice system with varying degrees of structural disorder to investigate how fluctuations in the material properties affect electron-hole separation. We also demonstrate how proper statistical treatment
of the CT energies recovers experimentally observed "hot" and "cold" exciton dissociation pathways.
Using a quantum mechanical model for a model heterojunction interface,
we recover experimental values for the open-circuit voltage at $50$ and $100 \ \rm{meV}$ of site-energy disorder.
We find that energetic and conformational disorder 
generally facilitates charge transfer; however, due to excess energy supplied by photoexcitation, highly energetic electron-hole pairs can dissociate in unfavorable directions, potentially never contributing to the photocurrent. We find that "cold" excitons follow the free energy curve defined at the operating temperature of the device.
Our results provide a unifying picture 
linking various proposed mechanisms for charge separation.
\end{abstract}

\section{Introduction}
\begin{figure}[t]
\centering
\includegraphics[scale=0.77]{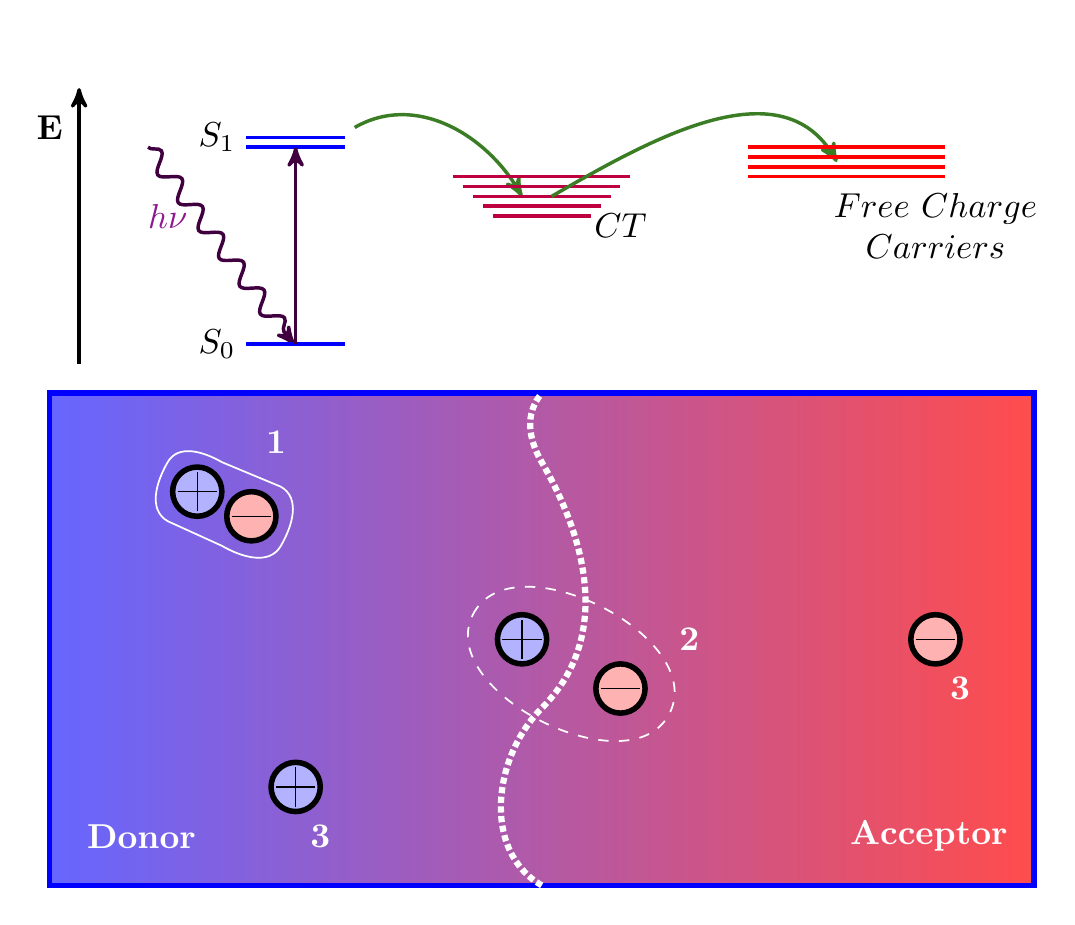}
\caption{{\bf Schematic diagram of exciton dissociation and formation of free charges in bulk-heterojunction OPVs.} Photoexcitation produces an exciton (1) in the donor,the acceptor or at the interface between them. Excitons created in the donor phase diffuse toward the interface (2) where energetic offset forces them to dissociate and separate apart, ideally, becoming free charges (3).}\label{fig1}
\end{figure}

Abundance of sunlight on the surface of the earth strongly favors solar cells as the replacement of fossil fuels as a primary energy resource. We are particularly interested in organic photovoltaic devices (OPVs) that have attracted considerable attention due to their promising electronic properties, cost effectiveness, and customizability. \cite{appenerg2017chatzisideris-krebs,natphot2009park-heeger,pps2013scharber-sariciftci, natphot2015cao-russel-he} However, the lack of complete understanding of physical processes that guide current generation in morphologically complex OPVs has stymied their full potential on a commercial scale. Figure~\ref{fig1} outlines the elementary steps of energy transfer and charge generation following photo-excitation in a donor-acceptor system

Successful dissociation of electron-hole pairs separated by the donor-acceptor boundary (charge-transfer (CT) states) into free charge carriers is especially puzzling since it requires an electron and a hole to overcome the strong electrostatic attraction under seemingly unfavorable conditions. More specifically, we consider an electron \textit{free} when the strength of the Coulombic interaction that binds it to the hole is comparable with thermal fluctuations of the material. This condition can be expressed in the following way

\begin{equation}
\frac{e^{2}}{4 \pi \epsilon _{0} \epsilon r} \leq k_BT,
\label{eq1}
\end{equation}

where $e$ is the electron charge, $\epsilon_{0}$ is vacuum permitivity, $\epsilon$ is a dielectric constant of the material and $r$ is the distance of separation between an electron in the conduction band a hole in the valence band. Due to the low dielectric constant ($\epsilon =2-4$) of organic materials typically used in OPVs, an electron finds itself with a 0.5~eV barrier to surmount, corresponding to Coulomb capture radius of 15 to 28~nm. It is highly improbable that charges would be able move this distance before recombining; however, it has been observed that free charge carriers can be formed at separations of 4~nm on the femtosecond timescales. \cite{science2014gelinas-rao-friend,jacs2016jakowetz-friend} 

The apparent disparity between the simple energetic estimate and the experimental evidence stems from the fact that entropic effects are neglected in the estimate provided by Equation~\ref{eq1}. Durant and Clarke have pointed out that an electron can take various possible paths to decouple from the hole and reach the free carrier state \cite{chemrev2010clarke-durant}. The number of possible dissociation routes increases the farther the electron travels as does the electron's density of states for a corresponding electron-hole separation and energy.  Consequently, energetic considerations alone do not offer complete picture of the charge dissociation and to gain better understanding of the process one should compute the (Helmholtz) free energy as a function of electron/hole separation distance, $r$.

\begin{equation}
F(r) = U(r) - TS(r) = U(r) - k_BTln\Omega(r)
\end{equation}
where $\Omega(r)$ is the the number of equivalent
electron/hole states with separation $r$, 
$T$ is the absolute temperature, $k_{B}$ is the Boltzmann constant, $U(r)$ is the electron-hole interaction potential, 
and $S$ is the entropy of the electronic degrees of freedom.

In an ordered material, dimensionality is the key factor that determines the relevance of the entropy term. Gregg argues that an $\pi-$electron confined to move along a single quasi-one-dimensional polymer chain has only one defined path and consequently the electronic entropy is exactly zero. Moreover, for thin-films (2-D) and fullerene-based acceptors (3-D) the number of electron/hole configurations available to the system with a given electron/hole separation radius scales with the surface area 
$$
\Omega \propto (r/r_o)^{d-1}.
$$
where $r_o$ is the unit length.
Consequently, the entropy term, $TS = T(d-1)\ln (r/r_o)$, can become energetically comparable to the Coulombic energy of the electron-hole pair in two and three dimensions. \cite{jpc2011gregg,jpcl2016hood-kassal} This estimate, however, is only valid for the scenarios with immobile hole. Allowing the hole to move adds additional degrees of freedom and increases the number of available electronic states further emphasizing the importance of entropic contribution. 

Generally speaking, organic semiconductors are disordered and amorphous systems.  Besides dimensionality, the free-energy needs to reflect a non-uniform landscape of donor and acceptor energies. This implies that neither $U$ nor $S$ are dependent upon the inter-exciton distance alone. For example, entropy reflects the number of paths that an electron and a hole can take as they separate apart; however, due to the finite life-time of the CT state, a sizable fraction of the possible dissociation routes becomes unavailable. This has to be included in the theoretical description. \cite{jpc2011gregg}

Entropy and free energy are state functions and to employ them we need to be certain that the system we are studying is in equilibrium. Burke {\em et al.} argue that this is the case for organic photo-excited materials since post-dissociation electron-hole encounters do not result in immediate recombination. Instead, electrons and holes meet and separate several times before, eventually, recombining. Faster rate of separation leads to a rapid equilibrium between interfacial charge-transfer states and free carrier species. This equilibrium condition then implies an equality between chemical potential and the open-circuit voltage, $qV_{OC}$ of the photo-voltaic device.\cite{advenmat2015burke-vandewal-mcgehee} 

This equality, therefore, establishes a crucial connection between theoretical and laboratory investigations of current generation in OPVs. Burke and colleagues arrive at the following  expression for the $V_{OC}$ from the canonical ensemble:
\begin{equation}
qV_{OC} = E_{CT} - \frac{\sigma_{CT}^{2}}{2kT} 
- kT\log \left( \frac{qfN_{0}L}{\tau_{CT}J_{sc}} \right)
\label{eq3}
\end{equation}
where $f$ is the volume fraction of the device that is mixed or interfacial, $L$ is the thickness of the solar cell,  $J_{SC}$ is the short-circuit current of the cell, $q$ is the electric charge, and $N_{0}$ is the density of the electronic states in the device. Most importantly, Equation~\ref{eq3} includes the necessary dependence of $qV_{OC}$, and therefore of $F$, on the average energy of the CT state, $E_{CT}$, disorder in the CT energies, expressed through standard deviation $\sigma_{CT}$, and the life-time of the CT-state, $\tau_{CT}$.\cite{advenmat2015burke-vandewal-mcgehee} Nonetheless, this expression is composed of variables that refer to the entire device making Equation~\ref{eq3} computationally inapplicable.

Following Clarke and Gregg's work, and investigating the effects of both dimensional entropy and energy disorder on the exciton dissociation, Hood and Kassal followed 
an approach based upon statistical mechanics
to compute the change in free energy, \cite{jpcl2016hood-kassal}
\begin{equation}
F = -\langle k_{B}T\ln Z\rangle \label{eq4}
\end{equation}
where $Z$ is the partition function that describes specific energy states. The bracket $\langle \cdots \rangle $ denotes statistical averaging over calculations done on the hexagonal lattice model representing the bulk-heterojunction. 
Both Equations~\ref{eq3} and~\ref{eq4} are derived in the canonical ensemble and carry the same information. 
However, the statistical approach is far more suitable for 
connecting to microscopic details such as energetic and 
structural disorder.

In the Hood-Kassal model, the donor and acceptor domains are modeled as a hexagonal lattice; however, electrons and holes are not permitted to cross from one domain to the other.  Furthermore, the hole is restricted to move only perpendicular to the interface due to translational symmetry. The electrostatic potential is the only interaction included in the model. They derive and equation based on the geometry of the lattice that describes density of states as a function of electron-hole separation and use it to compute entropy. The model does not take into account important quantum effects such as delocalization, mixing between excitonic and charge-transfer configuration, and electronic exchange effects.

In agreement with previous studies, Hood and Kassal have shown that change in free energy better reflects the energy landscape that an electron and a hole traverse, since both entropic considerations and energetic disorder tend to lower the energy barrier needed for an electron to become a free charge carrier. The energy of Coulombic interaction becomes comparable to thermal fluctuations and no longer defines how far an electron and a hole can separate. \cite{jpcl2016hood-kassal}

In this work, we use a fully quantum mechanical model of the electronic states of a bulk-heterojunction interface to investigate how the presence of entropy and disorder in OPVs influences the free energy of an electron as it separates away from the interface into the free carrier phase, taking into account Coulombic and exchange interactions, lattice vibrations, and electron-phonon couplings. In addition, we examine the role of band-width and interfacial driving forces in determining the dissociation free energy of an electron/hole pair. Our computationally cheap model combines quantum and statistical treatments of a system with large number of parameters and all possible electron-hole configurations to give results that provide a unifying picture linking various proposed mechanisms for charge separation.

\section{Methods}
\begin{figure}[t]
\hspace*{-0.5in}
\centering
\includegraphics[scale=0.5]{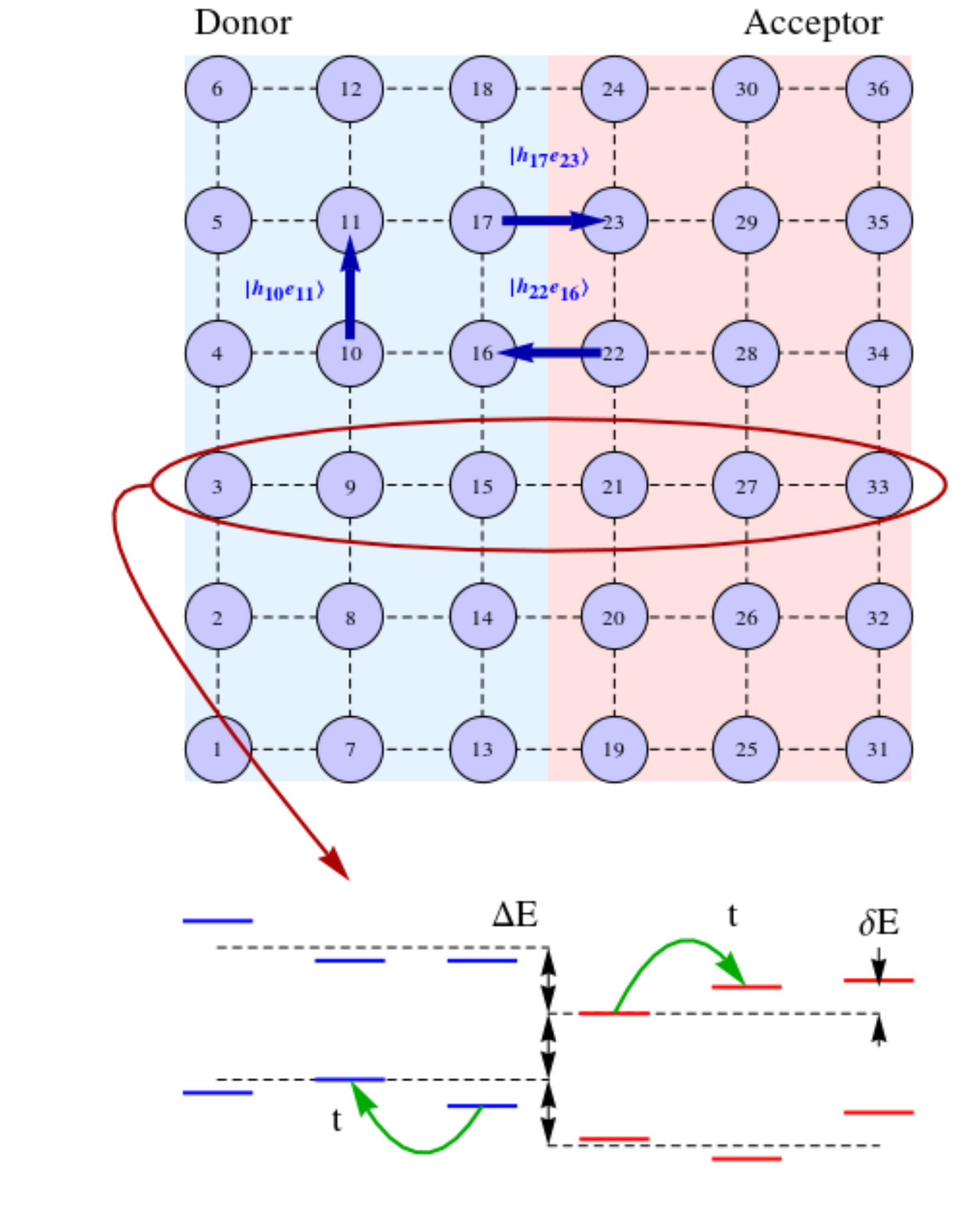}
\caption{{\bf Schematic representation of the lattice model and conduction and valence band energies associated with each site.} Donor and acceptor phases are distinguished by lowering acceptor site energies by $\Delta E$. $t$ is the hopping parameter controlling electron and hole's movement between neighboring sites. Disorder is introduced into the model by modifying energy of each site by some value randomly drawn from a Gaussian distribution with variance $\delta E$. $|h_{17}e_{23}\rangle$, $|h_{22}e_{16}\rangle$ and $|h_{10}e_{11}\rangle$ are examples of CT, "flipped" CT and an intraphase CT configurations respectively.}\label{fig2}
\end{figure}

The region of donor-acceptor interface of an OPV is generalized to the square lattice system, as shown in Figure~\ref{fig2}, where each site contributes a valence and a conduction band coupled to two phonon modes associated with lattice vibrations. Donor and acceptor sites are differentiated through the energetic offset in their band energies, $\Delta E$. The model is described by the following system-plus-bath Hamiltonian,\cite{EXCITON_MAIN,EXCITON_MAIN2}
\begin{equation}
\centering
\begin{split}
\hat{H} &= \hat{H}_{el} + \hat{H}_{el-ph} + \hat{H}_{ph} \\
&=\sum_{\textbf{mn}}(F_{\textbf{mn}}+V_{\textbf{mn}})|\textbf{m}\rangle\langle\textbf{n}|
+ \sum_{\textbf{mn},\mu}\Big(\frac{\partial F_{\textbf{mn}}}{\partial
  q_{\mu}}\Big)q_{\mu}|\textbf{m}\rangle\langle\textbf{n}| \\
  &+
\frac{1}{2}\sum_{\mu}\omega^{2}(q_{\mu}^{2}+\lambda q_{\mu,\mu+1}) + p_{\mu}^{2} 
\end{split}
\end{equation}
$\hat{H}_{el}$ represents electronic configuration interactions (CI) of singlet and triplet electron-hole states $|\textbf{n}\rangle$ described by localized Wannier functions.

We have published the details and parameterization of the model previously, here we briefly review its salient features. The single-body term $F_{\textbf{mn}}$ defines site energies of a valence band hole and a conduction band electron, as well as transfer integrals between neighboring sites. $V_{\textbf{mn}}$ describes spin-dependent two-electron interactions, namely Coulomb and exchange integrals for an electron and a hole occupying different sites, and dipole-dipole integrals for geminate singlet electron-hole pairs.\cite{EXCITON_MAIN2} Inclusion of exchange and dipole-dipole interactions in our model allows for a more realistic description of electrostatic attraction between an electron and a hole. While the model can treat singlet or triplet states, we are analyzing post-photoexcitation charge-transfer states and will only focus on singlets in this study. The phonon term, $\hat{H}_{ph}$, describes the vibrations of the lattice sites with two sets of local harmonic oscillators with weak nearest-neighbor coupling. Since OPVs often employ polymers, we consider the frequencies roughly corresponding to carbon-carbon double bond stretching and ring-torsional motions to be the most significant within this model. These vibrations modulate electron-hole energy levels and site-to-site transfer integrals. We include this modulation through the electron-phonon coupling term, $\hat{H}_{el-ph}$, calculated from the empirical values of the normal modes and corresponding Huang-Rhys parameters. 

We perform our calculations in an complete electron/hole configuration basis where each ket $|\textbf{n}\rangle$ represents a state with an electron on site $i$ and a hole on site $j$. Once we map the states onto a lattice in the Cartesian plane, we can compute electron-hole separation $R_{ij}=\sqrt{(x_{i}-x_{j})^{2}+(y_{i}-y_{j})^{2}}$, where $(x_{i},y_{i})$ and $(x_{j},y_{j})$ are site-wise coordinates of a hole and an electron respectively. We also define an operator $\hat{R}$ such that $\hat{R}|\textbf{n}\rangle = \hat{R}|\textbf{h}_{i}\textbf{e}_{j}\rangle= R_{ij}|\textbf{h}_{i} \textbf{e}_{j}\rangle$ to compute the expected e-h separation $\langle R \rangle$ for k-th eigenstate $|\Psi^{k} \rangle$ of $\hat{H}_{el}+\hat{H}_{el-ph}$ and match it to the corresponding eigenvalue. 

\begin{equation}
\langle R^{k} \rangle = \langle \Psi^{k}|\hat{R}|\Psi^{k} \rangle = \sum_{ij} |\rho^{k}_{ij}| R_{ij}
\end{equation}
where $|\rho^{k}_{ij}|$ represents contribution of configuration $|\textbf{h}_{i} \textbf{e}_{j}\rangle$ to the k-th eigenstate $|\Psi^{k}\rangle = \sum_{ij} c^{k}_{ij}|h_{i}e_{j}\rangle$. 

Removing any restriction on where on the lattice an electron and a hole can be located gives us a more detailed model of the system but comes at a disadvantage, because now, in addition, to charge-transfer states defined above, we need to consider ``flipped'' CT states with holes in the acceptor and electrons in the donor, as well as states where both particles are located in the same phase. To clarify the nature of each eigenstate we think of $\rho_{ij}$ as the charge of the dipole moment associated with configuration $|h_{i}e_{j}\rangle$ with $\rho_{ij}>0$ for $i \geq j$ and  $\rho_{ij}<0$ for $j>i$. We then compute the probability of $|\Psi^{k}\rangle$ having a CT character and its expected dipole size, $\langle R_{D} \rangle$.

\begin{equation}
\langle R^{k}_{D} \rangle = \frac{\langle \Psi|\hat{D}|\Psi \rangle}{q} = \frac{\sum_{ij} \rho^{k}_{ij} R_{ij}}{q}
\label{eq7}
\end{equation}

 Figure~\ref{fig2} illustrates some of the possible configurations of the system to explain the difference between $\langle R \rangle$ and $\langle R_{D} \rangle$. State $|h_{22}e_{16}\rangle$ has the same e-h separation as charge-transfer state $|h_{17}e_{23}\rangle$ while its dipole is reversed. In addition, it is important to realize that for each configuration $|\textbf{h}_{i}\textbf{e}_{j}\rangle$ there is a configuration $|\textbf{h}_{j}\textbf{e}_{i}\rangle$, however $|\rho_{ij}|$ does not have to equal $|\rho_{ji}|$. Configurations such as $|h_{i}e_{i}\rangle$ do not contribute to the polarization of the system because $R_{ii}=0$. Configuration $|h_{17}e_{23}\rangle$, $|h_{22}e_{16}\rangle$ and $|h_{10}e_{11}\rangle$ represent CT, "flipped" CT and an intraphase CT states respectively. We can visualize the relationship between the diabatic energies (the eigenvalues of $\hat{H}_{el}+\hat{H}_{el-ph}$) and $\langle R \rangle$ or $\langle R_{D} \rangle$ to determine what changes in model parameters facilitate formation of free charges. We believe that $\langle R_{D} \rangle$ gives the e-h separation, but comparison between $\langle R_{D} \rangle$ and $\langle R \rangle$ is essential to determine CT character of each state.

In this study we consider a $10\times 10$ lattice with band energies of the acceptor sites $0.5 \ {\rm eV}$ lower then those of the donor sites (energetic offset $\Delta E$), while the value of the transfer (hopping) integral is the same in both directions throughout the system ($t = 0.536 \ {\rm eV}$). We define the distance between two neighboring (horizontally or vertically) sites as $1 \ {\rm a}$ or one lattice spacing unit. 

The average energy, $\widetilde{E} = \sum_{\textit{i}}E_{\textit{i}}/N$, is computed for each interval $\langle R \rangle + \delta R$ ($\langle R_{D} \rangle + \delta R$) from $N$ energy levels that fall into those particular intervals. Similarly we compute the partition function and change in free energy:
\begin{eqnarray}
\Delta F &=& -kT\ln Z \nonumber \\ 
&=& -kT\ln\Big(Tr[e^{-\beta (\hat{H}_{el}+\hat{H}_{el-ph})}]\Big) \\
&=& -kT\ln\Big(\sum_{j} e^{-\beta E_{j}}\Big)
\label{eq8}
\end{eqnarray}

\section{Results and Discussion}
Disorder present in OPVs introduces fluctuations into the values of site parameters and therefore increases the number of states and energies available to the system. We create 100 realizations of the lattice system, with band energies of each site randomly drawn from a Gaussian distribution. The mean of said distribution is equal to the ordered site energy and variance, $\delta E$, represents the amount of disorder in the material. Each realization is a set of possible states of the system and we obtain fuller energy-separation and energy-dipole spectra by combining eigenvalues from all of the realizations. We first consider the system with low site-energy disorder ($\delta E = 10 \ \rm{meV}$). 

\begin{figure}[h]
\centering
\includegraphics[scale=0.15]{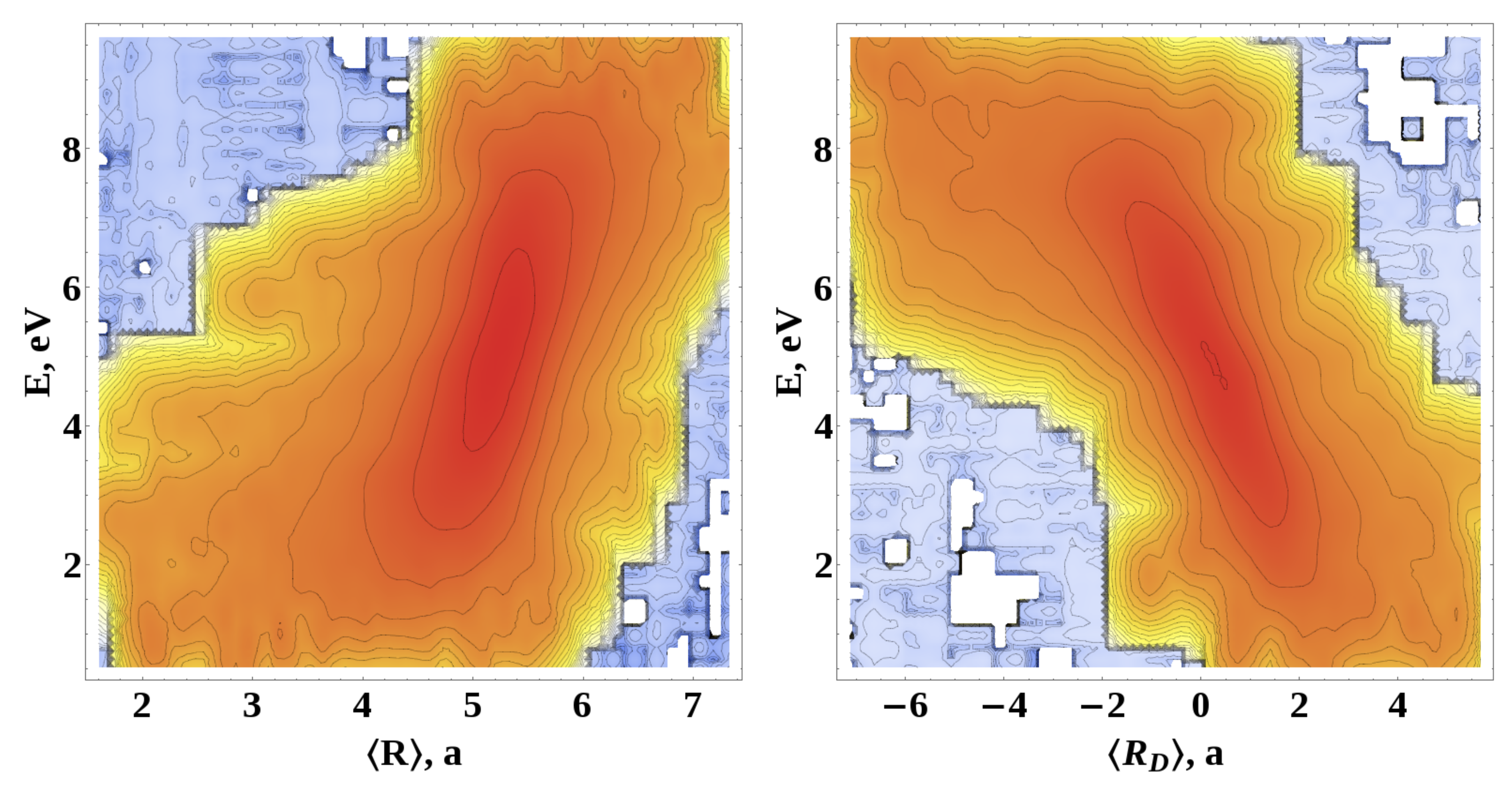}
\caption{{\bf Distributions of CT energies of the system as a function of electron-hole separation, $R$, and dipole size, $R_{D}$.} The color map is scaled such that blue and red correspond to the lowest and highest densities respectively. The sign of the dipole size depends on the orientation of the CT dipole: if an electron is in the acceptor phase and a hole is in the donor the dipole size is positive, otherwise it takes on negative values. Distributions are computed for systems with $\delta E = 10 \ {\rm meV}$}
\label{fig3}
\end{figure}

Figure~\ref{fig3}a shows a very broad distribution of \textit{e-h} energies and separations, with average energy of the electron-hole pair increasing as charges move farther apart. Figure~\ref{fig3}b adds important detail to identifying the nature of the states by introducing a general sense of the state's direction through the dipole moment. Negative value of $\langle R_{D} \rangle$ indicates that configurations pointing from the acceptor to the donor contribute the most to a particular state of the system. The densest region of both distributions corresponds to the inversion point, where an electron and a hole of the charge-transfer state trade places. In Figure~\ref{fig3}a this point is accompanied by the increase in density of states around $(5.5 \ {\rm a},5 \ {\rm eV})$ and is followed by the sharp raise of energy for larger values of $\langle R \rangle$. We can interpret this as additional kinetic energy allowing an electron and a hole to traverse the lattice independently of the phases. However, it is also an artifact of using a system of a finite size. When a hole resides at the interface, the maximum distance an electron can separate is limited by the length of the acceptor region. As energy increases, more and more states that have achieved farthest possible separation accumulate increasing the density of states at $\langle R \rangle \approx 5.5 \ {\rm a}$. Eventually, \textit{e-h} pair has enough energy to transfer to the state where a hole can move into the acceptor and an electron moves into the donor to maximize the distance between them. While this computational artifact would be greatly reduced by the simulation of a larger lattice, it is still important to consider since constrained acceptor regions do occur in the morphologically complex OPVs.

\begin{figure*}[ht]
\centering
\includegraphics[scale=0.15]{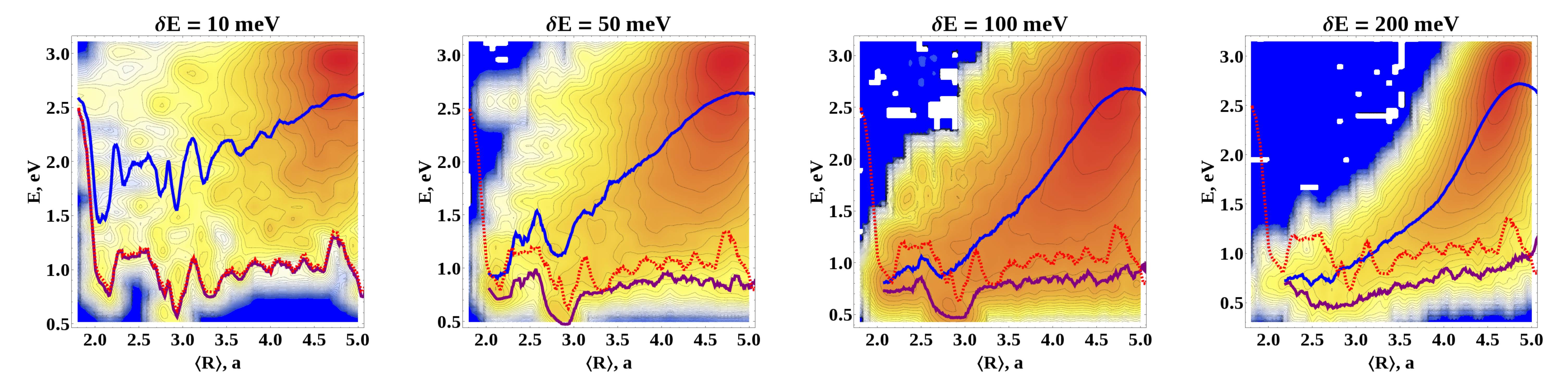}
\includegraphics[scale=0.15]{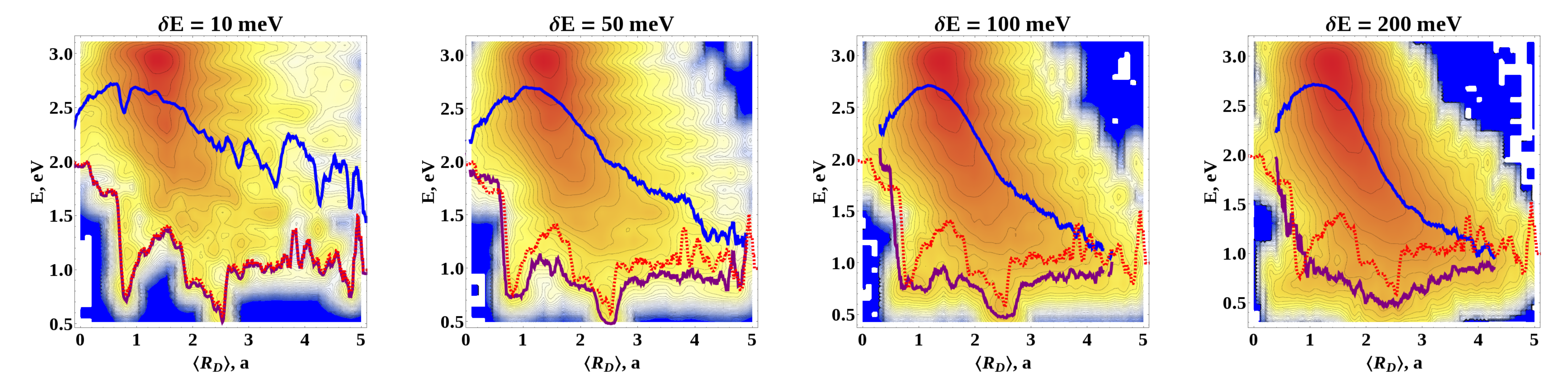}
\caption{{\bf Effect of site-energy disorder on the state energy distributions and average and free energies of the system.}  Each panel shows charge distribution computed for specific values of site-energetic disorder $\delta E$. The color map is the same as in Figure~\ref{fig2}. Blue curve represents average energy as a function of electron-hole separation, $\widetilde{E}(R)$; dashed red curve is thermodynamic energy, $\langle E(R) \rangle$; purple curve - free energy, $F(R)$ White islands in each plot are plotting artifacts and do not carry any information.}\label{fig4}
\end{figure*}

Bittner and Silva previously showed using this model that site-energy fluctuations couple photoexcitations to the well-separated CT states allowing hot excitons to be converted into photocurrent within $100 \ \rm{fs}$. \cite{natcomm2014bittner-silva} The primary objective of this paper is to investigate the effects of energetic disorder on the the average and free energies of the system and consequently on the \textit{e-h} separation.

The full CI approach for the $10$ by $10$ system results in 10,000 eigenstates for each realization with 90\% of theses states' eigenvalues corresponding to the initial photons in the ultraviolet region. For the rest of the paper we will focus on states of the system corresponding to the energies less then or equal to $3.1\ {\rm eV}$. To avoid issues with the lattice size we consider the separations up to $5~{\rm a}$. It is also worth pointing out that the system is very sensitive to the presence of the disorder: even $\delta E = 10~{\rm meV}$ is approximately 4 times greater then the energy difference between the two consecutive CT states.

Figures~\ref{fig4}a and \ref{fig4}b demonstrate how more energetic regions of each distribution become smoother and narrower as the amount of disorder in the system increases. Average energy curve, $\widetilde{E}(\langle R \rangle)$, shows that as the value of $\delta E$ grows energetic barriers obstructing charge separation disappear but overall energy needed to move an electron and a hole apart increases from $1 \ \rm{eV}$ to almost $2 \ \rm{eV}$. At the same time, in Figure 4b greater disorder pushes more energetic states toward smaller dipole sizes indicating that energy in access of $1.5 \ {\rm eV}$ facilitates charge separation not only in CT but does so for interphase and "flipped" CT configurations. Larger values of $\langle R_{D} \rangle$ with energies below $1.5 \ {\rm eV}$ predominantly correspond to CT states with an electron in the acceptor and a hole in the donor, however as energy of the state increases so do the contributions from other far separated configurations reducing the size of the dipole. Distributions in both Figure 4a and 4b show that increasing disorder broadens the low energy regions by creating new low energy state.

Average energy curves $\widetilde{E}(\langle R \rangle)$ and $\widetilde{E}(\langle R_{D} \rangle)$ connect states that are most likely to be observed at a given $\langle R \rangle$ or $\langle R_{D} \rangle$ and represent the energetic path that an exciton takes to dissociate into free charge carriers only if all of the states are equally probable. Diagonalizing the Hamiltonian returns only the values of all possible energy levels, and it is the operating temperature of the device that determines the probability weight of each state. OPVs are in thermal equilibrium with the environment and their thermodynamic energy for a specific \textit{e-h} separation should be computed as follows:\cite{tuckerman2010statmech}
\begin{equation}
\langle E \rangle = \frac{\sum_{j} E_{j}e^{-\beta E_{j}}}{\sum_{j} e^{-\beta E_{j}}}\label{eq9}
\end{equation}
where $\beta = {1}/{k_{B}T}$, $E_{j}$ is j-th eigenstate of the system at a certain $\langle R \rangle$ ($\langle R_{D} \rangle$) and $e^{-\beta E_{j}}$ is its thermal probability weight. We consider realizations of the system to form a canonical ensemble, since there is no particle exchange between the device and its surroundings. We compute $\langle E \rangle$ for the reference system at $T = 298 \ {\rm K}$ and show that at this temperature thermal fluctuations are not strong enough to support more energetic CT states. Since $e^{-\beta E_{1}}>>e^{-\beta E_{N}}$, where $N$ represent states with $E_{j} > 1.5 \ {\rm eV}$, the lowest energy states become strongly favorable and dominate both sums of the Equation~\ref{eq9}. The red dashed curve in Figure 4 therefore traces out the most likely path an exciton will take toward generation of free carriers in the system with no disorder at room temperature.

The free energy computed in Equation~\ref{eq7} through the partition function takes into the account thermal probability weight of each state as well as the change in entropy of the system due to the presence of disorder. We focus on the structural disorder because thermal fluctuations would not be able to sustain charge-transfer states and previous studies relate disorder in electronic parameters of the OPVs with structural fluctuations in the material. \cite{jpcl2017simine-rossky,pccp2015bassler-kohler}. For the ordered system at $T = 298 \ {\rm K}$, $F \approx \langle E \rangle$ and we use it as a reference when studying the effects of increased disorder. Our simulations show that increasing the value of site-energy disorder even to $50 \ {\rm meV}$ drastically reduces the height of energy barriers that could obstruct charge generation. When the amount of disorder in the system is $100~{\rm meV}$, fluctuations completely remove most of the major and minor peaks from the free energy curve. We have noted previously that in the case of $\delta E = 10 \ {\rm meV}$ the distribution of CT states with energies less then $1.5 \ {\rm eV}$ is discontinuous and as the value of $\delta E$ increases these gaps in this distribution fill up with new states. While the number of CT states with energies above $2 \ {\rm eV}$ is far greater then the number of newly generated lower energy states, these low energy states are far more likely to be occupied because at $298 \ {\rm K}$ they are thermally preferable. Presence of disorder reduces electrostatic attraction between an electron and a hole by creating additional states forming a more efficient dissociation path. In the case of average energy, $\widetilde{E}$, additional lower energy states thus smooth out and decrease $\widetilde{E}(R)$ for $R<4.5 \ {\rm a}$ and $\widetilde{E}(R_{D})$ for $R_{D}>1.5 \ {\rm a}$, but do not have a significant effect at the dense regions of the distributions. 

\begin{figure*}[ht]
\includegraphics[width=0.48\textwidth]{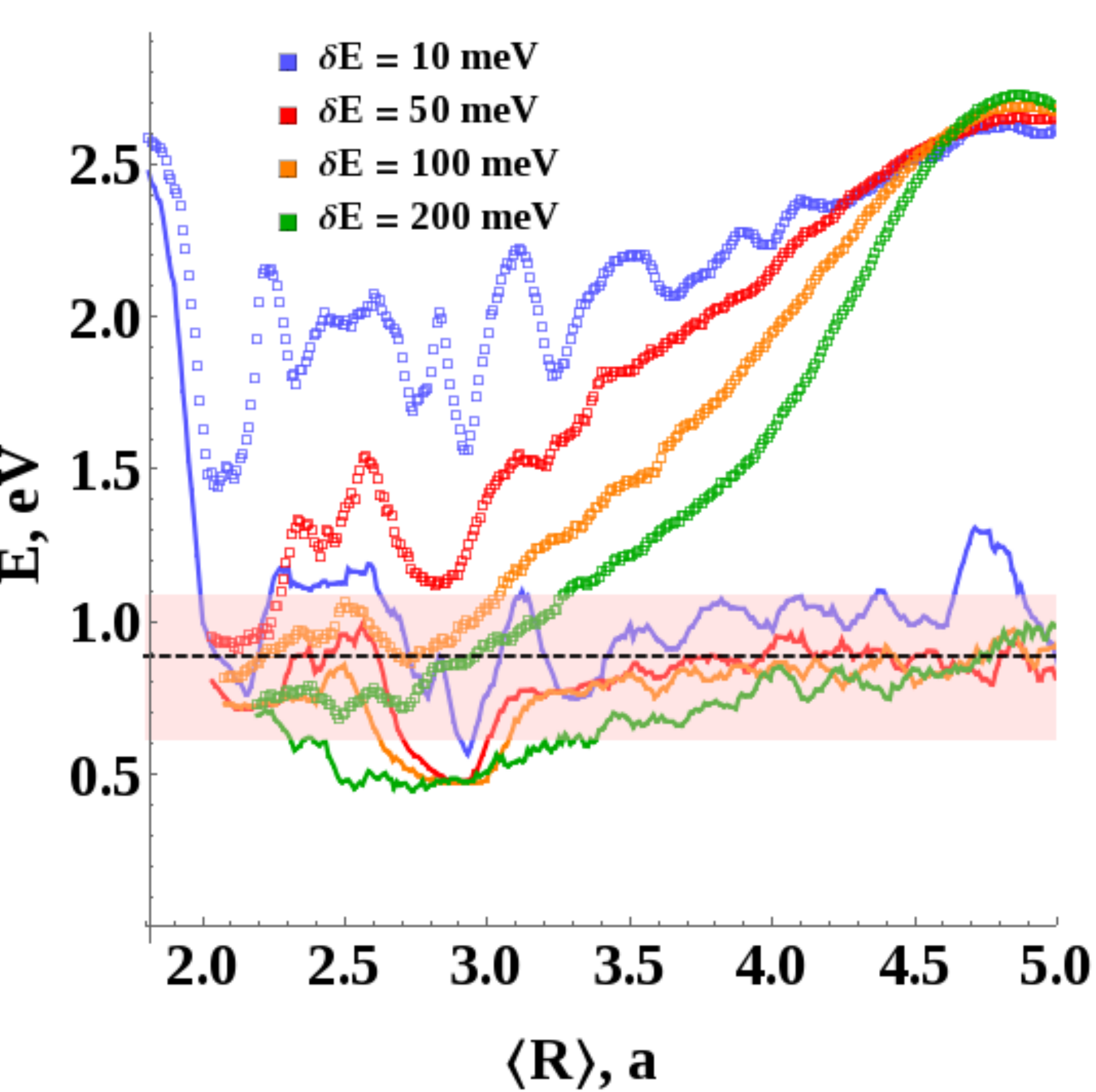}
\includegraphics[width=0.48\textwidth]{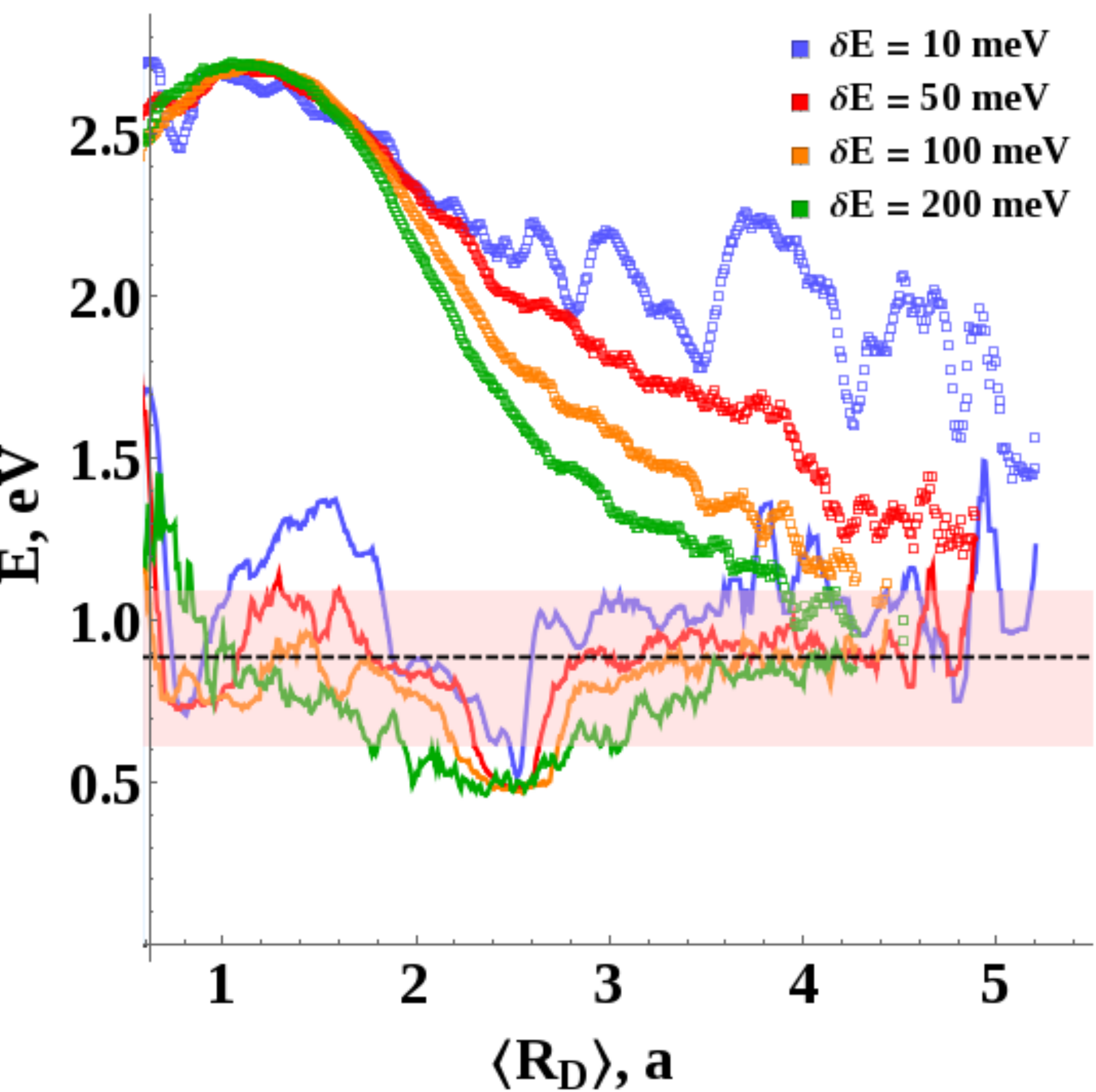}
\caption{{\bf Average (top four) and free energy (bottom four) curves at different values of site-energy disorder $\delta E$.} Shaded pink region represents the range of experimental values for open-circuit voltage with the average indicated by the dashed black line.}
\label{fig5}
\end{figure*}

\begin{figure*}[t!]
\begin{subfigure}[]
\centering
\includegraphics[scale=0.4]{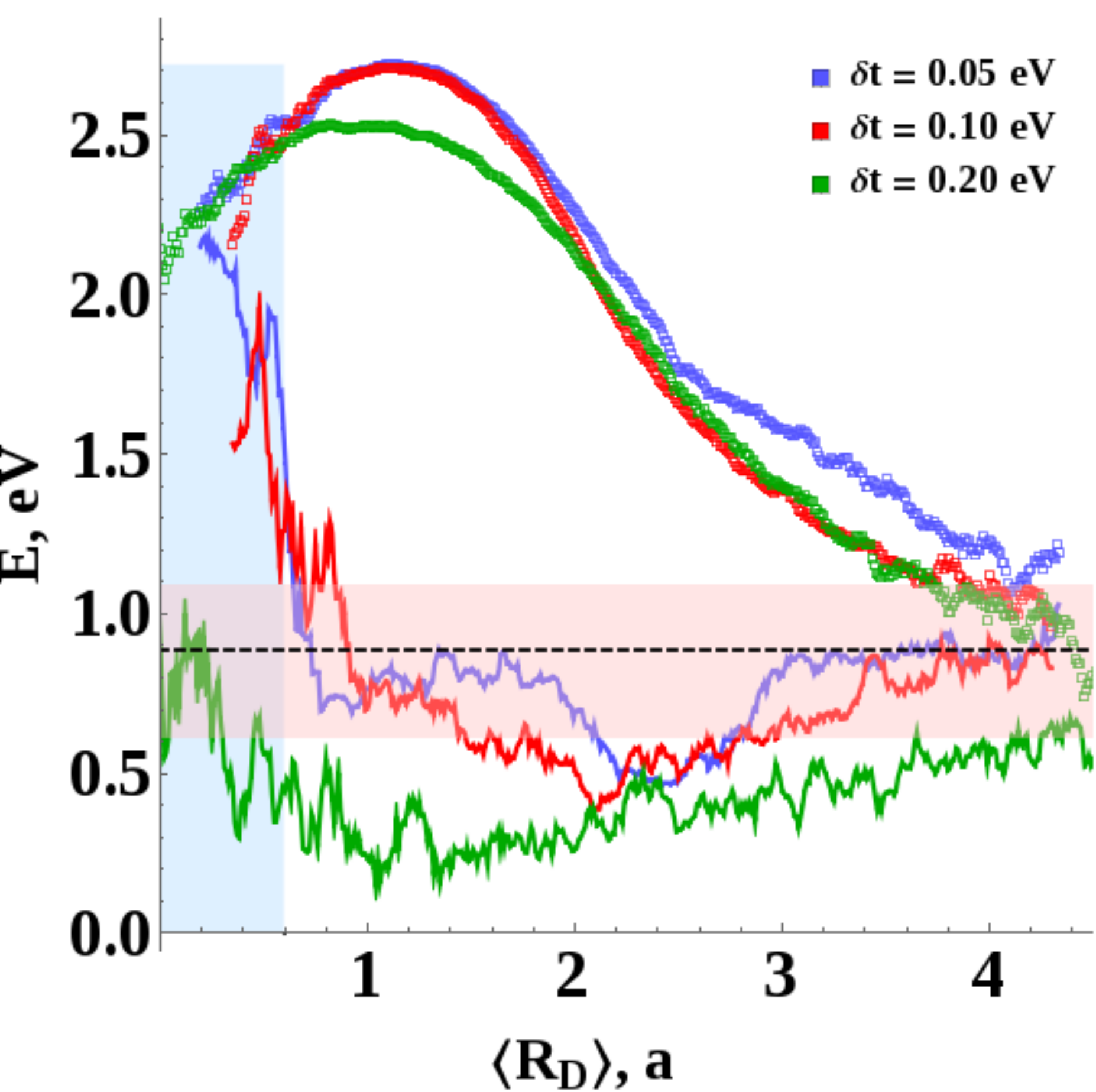}
\label{a}
\end{subfigure}
\hspace*{0.2in}
\begin{subfigure}[]
\centering
\includegraphics[scale=0.4]{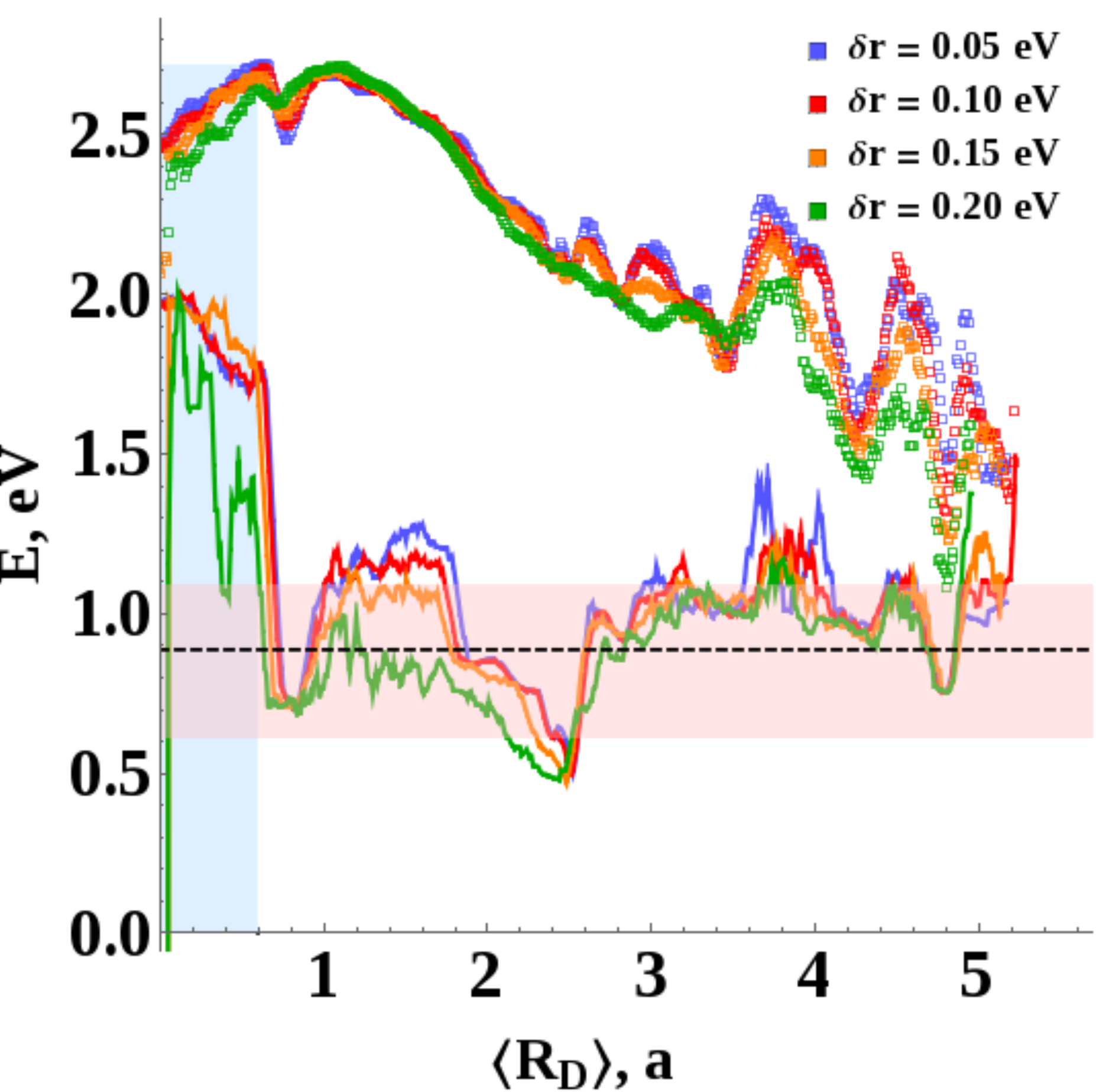}
\label{b}
\end{subfigure}
\begin{subfigure}[]
\centering
\hspace*{-0.5in}
\includegraphics[scale=0.15]{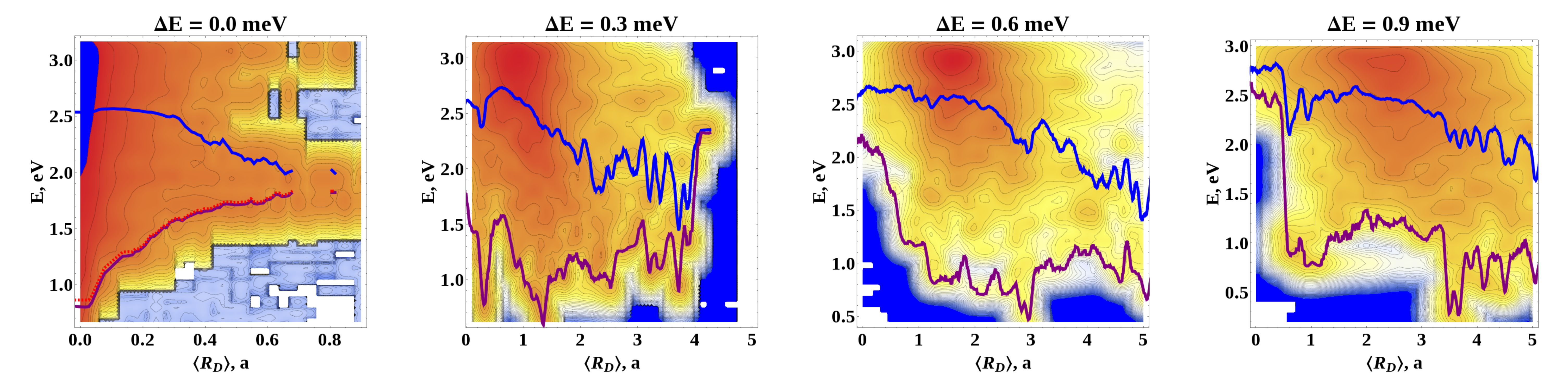}
\label{c}
\end{subfigure}
\caption{{\bf The effect of hopping and configurational disorder on the electron-hole separation.} Average and free energies for different amounts of a). hopping disorder, $\delta t$, and b). disorder in site coordinates, $\delta r$, compared against experimental results as in Figure 6. c). Distribution of CT energies and electron-hole separations and the corresponding average (blue) and free energies (red) plotted from simulations with variable donor-acceptor energetic offset, $\Delta E$}
\label{fig6}
\end{figure*}


We argue that the graphs of average ($\widetilde{E}(R)$) and free energy ($F$) curves illustrate the two experimentally observed mechanisms of charge separation: direct ultra-fast charge transfer and slower charge dissociation mediated by intermediate CT states.\cite{orgoelec2012sheng-vardeny} There is no consensus in literature regarding which pathway contributes the most to device performance. Investigations carried by Provencher \textit{et al.} indicate that "hot" excitons can become charge carriers right away without relaxation and lead to very high quantum efficiencies, while the effectiveness of electron transfer through the lower states heavily depends the type of the blend used and the morphology of the device.\cite{natcomm2014provencher-silva} Savoie \textit{et al.} showed that rapid charge dissociation of the "hot" exciton is facilitated by the high density of acceptor states. \cite{jacs2014savoie-ratner} On the other hand, Vandewal \textit{et al.} observed that internal quantum efficiency in various photovoltaic blends is independent of whether the initial excited state is $CT_{1}$ (lowest charge-transfer state) or any higher laying CT state. Measurements indicate that the very energetic states contribute little to the generation of free charges, because they relax within the CT manifold before they can achieve greater separations. \cite{natmat2014vandewal-salleo}. Our model recovers both pathways and allows us to explain a number of experimental results. 

$\widetilde{E}$ describes thermodynamic energy of the system in the high temperature limit ($\beta \rightarrow 0$). While the system cannot place an exciton in the more energetic CT state at $T = 298 \ {\rm K}$, the photons illuminating the device have sufficient amount of energy (for visible light $h\nu = 1.7 - 3.0~\rm{eV}$) to realize any of the possible states of the system. After the photoexcitation, however, the "hot" electron-hole pair finds itself in the system equilibrated at $298 \ {\rm K}$ where $F$ describes the energies of most probable states. Therefore, if the exciton does not dissociate into free charge carriers fast enough, it will relax to a lower CT state on the $F$ curve and follow that path toward the polaron state.

Figure~\ref{fig5} compares the average and the free energy curves at different values of $\delta E$ to see the effect of disorder on the charge separation. Since we consider all states equally probable in the high temperature regime, regions where density of states is higher contribute more toward $\widetilde{E}$. At the same time the greater number of states in the said regions mean that any energetic fluctuations would be averaged out. Figure~\ref{fig5}b shows that increasing site-energy fluctuations does not affect the average energy when $R_{D}<2 \ {\rm a}$ (corresponds to the densest part of the spectrum) and smooths and steepens the curve for greater separations favoring fast dissociation. At the same time, energies of the lowest states dominate free energy. Even though these states are not as numerous, their contribution, when thermally weighted, is far greater and any shift in their energy will be immediately reflected on the graph of $F$, making free energy much more responsive to the composition of the device. 

Rapid decrease of the average energy $\widetilde{E}(R_{D})$ in Figure~\ref{fig5}b and increase in the corresponding dipole size show that the excess energy exciton receives from the photoexcitation allows it to quickly overcome Coulombic attraction. Average energy as the function of absolute electron-hole separation (Figure~\ref{fig5}a) supports this observation as well but specifies that while an electron and a hole can separate to greater distances in the high temperature regime, there is no control over which phase the free charges will migrate to. As was mentioned earlier, all possible electron-hole configurations contribute to the wavefunction of each state and as the energy of the exciton increases, so do the interphase and "flipped" CT contributions. In the absence of external field "hot" excitons may dissociate into free charge carriers that will never reach electrodes. At the same time, the magnitude of the free energy in both Figures 5a and 5b does not drastically change with increasing $R$ and $R_{D}$ and large dipole sizes indicate that CT configurations are most significant for the corresponding states. We explain this prevalence of a particular configuration orientation by the presence of an internal potential difference, $V_{OC}$, that drives charges apart. In the introduction section of this paper we talk about the equality of the free energy and the open-circuit voltage of the device ($qV_{OC}$ for correct units.) The pink region of Figure 5 corresponds to the range of experimental values of the open-circuit voltage obtained from different types of OPVs with disorder ranging from $60$ to $104 \ {\rm meV}$.\cite{advenmat2015burke-vandewal-mcgehee} We can see that $F$ curves for $\delta E = 50$ and $100 \ {\rm meV}$ agree with the experimental data very well. We believe this agreement to be a crucial result not only because it validates the importance of the lower CT dissociation pathway, but  also because it establishes a strong connection between theoretical models and experimental findings. Looking at the results of our calculations from the perspective of open circuit voltage we note that too much disorder will negatively impact device performance. While the barriers inhibiting charge transfer completely disappear, free energy curves decrease below the experimental range, leading to the drop of $V_{OC}$ and efficiency. $V_{OC}$ is one of the key parameters determining the efficiency of a solar cell and if we can directly calculate it our experimentally tested model, then we can also study how efficiency of the device depends on the properties that are not as easily accessible in the laboratory setting. 

To better understand all aspects of disorder in OPVs we perform simulations of systems where we randomly modify values of the hopping integral ($\delta t$) and the site coordinates ($\delta r$). We start with the same reference system as for the $\delta E$ calculations. Results from the hopping parameter (Figure~\ref{fig6}a) show that even small values of $\delta t$ have very noticeable effect on lowering the free energy, comparable to the site-energy disorder of $200 \ {\rm meV}$. This is expected, because even $\delta t = 0.05 \ {\rm eV}$ is a significant fraction of the value of the hopping parameter ($t= 0.536 \ {\rm eV}$). Increase in the hopping disorder facilitates the movement of an electron and a hole through the lattice along the low CT pathway allowing even very low energetic states to reach higher separations. We suspect that fluctuations in the hopping parameter have little effect on the $\widetilde{E}(R_{D})$, because "hot" excitons have sufficient kinetic energy to traverse the lattice with ease regardless of the values of the hopping term. As in the case of large site-energy disorder, the device performace takes a hit when the free energy and therefore the open-circuit voltage decrease significantly for $\delta t \geq 0.05 \ \rm{eV}$. However, given that $0.05 \ \rm{eV}$ is already 10\% of the hopping value, larger amount of disorder does not seam realistic. Variation in site locations (Figure~\ref{fig6}b) similarly has little effect in the high temperature limit, but significantly reduces the free energy for the states with smaller dipole size. The reduced barrier is most likely due to the intraphase Coulombic attraction that is much more sensitive to changes in the electron-hole separation when the charges are closer together and are not phase separated.  

  We conclude this section with results from the simulations of systems with the variable energetic offset presented in Figure~\ref{fig5}c. $\Delta E$ is closely related to the open-circuit voltage of the OPV and promotes charge dissociation at the interface helping an electron and a hole to overcome electrostatic binding, but whether it has the desired effect is still debated. \cite{jpcl2015jackson-ratner,advmat2006scharber-brabec} Our calculations show that increasing the energetic offset between two phases clearly extends the maximum possible electron-hole separation from $0.6 \ \rm{a} - 0.8 \ \rm{a}$ at no offset to more then $4 \ \rm{a}$ at $\Delta E \ = \ 0.9 \ {\rm eV}$. We also show that unlike other studied parameters, greater values of offset can shift the entire distribution toward larger values of $R_{D}$. In the absence of energetic offset increasing site-energy disorder lowers Coulombic interaction and free energy of the system but barely extends overall charge separation, indicating that non-zero $\Delta E$ is necessary for the efficient charge dissociation to take place. Increasing energy offset to $0.9 \ \rm{eV}$ removes low energy states from smaller values of the dipole size and makes the average energy curve more gradual. In this case the electron-hole pair separated across the interface has more then enough energy to separate farther, but as with the other extreme disorder parameters, at the expense of the device performance. 

\section{Conclusion}
We use the quantum mechanical lattice model \cite{EXCITON_MAIN,EXCITON_MAIN2} to provide critical insight into the nature of entropy in OPVs and to show that the correct statistical treatment of the computed charge transfer states elucidates the origins of two prominent charge separation pathways observed in these materials. \cite{jpcl2016hood-kassal,science2012bakulin-friend,scirep2016vella-silva-bittner,jacs2009pensack-asbury,natmat2014vandewal-salleo,jpcl2017athanasopoulos-koehler}

Photovoltaic devices are in thermal equilibrium with the environment and the temperature of the OPV determines the correct probability of each CT state. At room temperature, the exciton only can dissociate through the lowest CT states. However, a photoexcitation of the material delivers sufficient amount of energy for any state to become equally probable, essentially mimicking the high temperature regime and as a result unlocking a completely different charge separation pathway composed of highly energetic CT states. Energetic pathways computed at $298  \ \rm{K}$ and in the "high" temperature case describe electronically "cold" and "hot" CT dissociation mechanisms observed and actively discussed in the literature. Our calculations show that "hot" excitons face steep energy gradient that favors fast dissociation. Increasing disorder smooths the average energy curve connecting corresponding CT states as well makes the curve steeper indicating that the excess energy supplied by the light allows e-h pair to overcome Coulombic attraction and any other obstructions of free carrier generation. If the exciton does not undergo ultrafast separation, it will relax and follow CT states that are thermally available at the temperature of the OPV device. To account for for the thermal probabilities of each state and also for the effects of disorder in the material we compute free energy of the system. Considering all possible electron-hole configurations shows that even though highly energetic charges may be well separated, they may not be able to reach electrodes and contribute to photocurrent since they can dissociate into any phase. Free energy acts as the driver of the excitations in the lower states, directing electrons into the acceptor phase of the OPV. However, since density of high energy CT states is significantly greater at higher energies, free energy at lower temperatures is much more sensitive to any changes in the material composition. 

Structural disorder can manifest itself through many different parameters of the system such as site energies, values of the site-to-site hopping integral, and sites' physical location. While the parameters are interconnected, we study them separately to estimate their individual effects on charge dissociation. We show that structural disorder creates additional low energy CT states, therefore creating new favorable dissociation pathways. These new states drastically reduce the height of the electrostatic energy barrier which explains how bound electrons and holes separate from each other so easily. 

Our calculations of the free energy for systems with 50 to 100~meV of site-energy disorder are consistent with values of $qV_{OC}$ measured in OPV devices with similar amounts of disorder.\cite{prb2010vandewal-manca,advenmat2015burke-vandewal-mcgehee} This result is particularly important since it strengthens the link between our model and empirical evidence and it establishes $V_{OC}$, and therefore $F$, as a crucial benchmark for theoretical models in general. Our simulations show that increasing site energy disorder above 100~meV shifts the entire free energy curve outside of the experimental range. While lower free energy means that an electron and a hole can separate effortlessly, it also implies lower open-circuit voltage and therefore lower efficiency. We find that the free energy of the system is very sensitive to changes in the site-to-site hopping parameters with even small variations in hopping energy pushing free energy below the experimental range. Configurational disorder lowers the height of the interphase-Coulombic barrier. We suspect that increasing site-energy disorder blurs the energetic difference between donor and acceptor regions, while variations in site location only affect the distance dependent electrostatic interactions. Additionally, the presence of an energetic offset between donor and acceptor phases extends the maximum electron-hole separation.

In this paper we focused upon the influence of energetic and configurational disorder on the device properties of OPV systems. Computational cost-effectiveness and the general transferable set of parameters makes our model a suitable tool to investigate the effects of molecular orientation, morphology and dimensionality whose importance has been greatly emphasized in the literature. \cite{jacs2016jakowetz-friend,chemmat2015sulas-ginger,natmat2013noriega-stingelin-vandewal,jpcl2015jackson-ratner,natphot2014tumbleston-ade,advfuncmat2009mayer-mcgehee} A proper and self-consistent statistical treatment of the CT states leads to the correct understanding of the process of charge separation.  By equating the free energy and the open-circuit voltage of the device, we establish a crucial connection between model predictions and experimental findings.

\section*{Acknowledgements}
The work at the University of Houston was funded in
part by the  National Science Foundation (CHE-1664971, MRI-1531814)
and the Robert A. Welch Foundation (E-1337). 

\section*{Author contributions}
E.R.B. conceived the project. V.L.
performed the calculations and 
analysis.  Both authors contributed 
to the drafting and editing of the 
manuscript

\section*{Additional information}
\subsection*{Competing financial interests}
The authors declare no competing financial interests.

\providecommand*\mcitethebibliography{\thebibliography}
\csname @ifundefined\endcsname{endmcitethebibliography}
  {\let\endmcitethebibliography\endthebibliography}{}

\end{document}